\documentclass[a4paper,11pt]{article}
\usepackage[inline]{enumitem}
\usepackage[multidot]{grffile} 
\usepackage{graphicx}
\usepackage{subfigure}
\usepackage[margin=1in]{geometry}
\usepackage{dcolumn}
\usepackage{bm}
\usepackage{amsmath}
\usepackage{amsthm}
\usepackage{framed}
\usepackage{soul}
\usepackage{amsfonts}
\usepackage{bbold}
\usepackage{dsfont}
\usepackage{amssymb}
\usepackage{color}
\usepackage{latexsym}
\usepackage{stackrel}
\usepackage{slashed} 
\usepackage{empheq}
\usepackage{fancybox}
\usepackage{pstricks}
\usepackage{indentfirst}
\usepackage{mathrsfs}
\usepackage{tablefootnote}
\usepackage{longtable}
\usepackage{multirow}
\usepackage{epsfig,psfrag}
\usepackage{subfigure}
\usepackage{mathtools}
\usepackage{setspace} 
\usepackage[utf8]{inputenc} 
\usepackage[scientific-notation=true]{siunitx} 
\usepackage{verbatim}
\usepackage{comment}
\usepackage[many]{tcolorbox} 
\usepackage{JHEPpub}
\usepackage{JHEPorcid} 
\usepackage[T1]{fontenc} 
\newcommand{\bk}{{\boldsymbol k}}

\newcounter{subfiggroup}
\newcommand{\savedthesubfigure}{}

\graphicspath{fig/}
\title{QED radiative corrections in inverse beta decay from virtual pions}
\author[a]{Oleksandr Tomalak\orcidlink{0000-0002-4827-5842}}
\emailAdd{tomalak@itp.ac.cn}
\affiliation[a]{Institute of Theoretical Physics, Chinese Academy of Sciences, Beijing 100190, China}
\abstract{Inverse beta decay (IBD), $\overline{\nu}_e p \to e^+ n \left( \gamma \right)$, is the main detection channel for reactor and supernova antineutrinos. To provide precise IBD cross sections at antineutrino energies $E_{\overline{\nu}_e} \gtrsim 10~\mathrm{MeV}$, we evaluate radiative corrections from virtual pions within the framework of heavy baryon chiral perturbation theory. At leading order, only the pion isospin-breaking contributions are not suppressed by the electron mass. At next-to-leading order, besides recoil effects, only the Wilson coefficient $c_4$ contributes to the kinematic dependence. However, its precise value is not relevant for IBD at relatively low energies since all next-to-leading order radiative corrections are relatively small. We find the kinematic dependence of the pion-induced QED radiative corrections at the level and below the uncertainty from the momentum dependence of the nucleon form factors. Our results enable sub-permille theoretical precision of charged-current elastic (anti)neutrino-nucleon scattering at (anti)neutrino energies $E_{{\nu}} \gtrsim 10~\mathrm{MeV}$.}

\keywords{Inverse beta decay, neutrino cross sections, effective field theory, electroweak radiative corrections, supernova (anti)neutrinos}

\begin{document}
\maketitle
%

\section{Introduction}

The first detection of antineutrinos was performed by the inverse beta decay (IBD) reaction $\overline{\nu}_e p \to e^+ n \left( \gamma \right)$ around 70 years ago~\cite{Cowan:1956rrn}. Advantageously, the prompt-delayed time-energy correlations between the prompt photon from positron annihilation on atomic electrons and the delayed photon induced by neutron capture allow for efficient background discrimination. IBD also plays a decisive role in the production and detection of supernova antineutrinos~\cite{Kamiokande-II:1987idp,Bionta:1987qt,IMB:1988suc,Hirata:1988ad}. Today, IBD is the main detection channel in reactor antineutrino experiments~\cite{KamLAND:2002uet,KamLAND:2004mhv,DayaBay:2012yjv,DoubleChooz:2012gmf,RENO:2012mkc,KamLAND:2013rgu,JUNO:2015zny,JUNO:2015sjr,DayaBay:2017jkb,DayaBay:2018yms,RENO:2018dro,DoubleChooz:2019qbj,JUNO:2020ijm,JUNO:2021vlw,JUNO:2022mxj,JUNO:2022lpc,JUNO:2024jaw,JUNO:2025gmd,JUNO:2025fpc}, studies of geoneutrinos~\cite{KamLAND:2011ayp,Borexino:2010dli,Borexino:2015ucj}, searches for supernova explosions~\cite{Super-Kamiokande:2007zsl,JUNO:2015zny,JUNO:2015sjr,KamLAND:2015dbn,Super-Kamiokande:2016kji,KamLAND:2024uia,JUNO:2021vlw} and the diffuse supernova background~\cite{Super-Kamiokande:2002hei,Super-Kamiokande:2002hei,Super-Kamiokande:2013ufi,JUNO:2015zny,Super-Kamiokande:2021jaq,JUNO:2021vlw}.

The IBD cross sections relate the antineutrino fluxes to the observed event rates. The cross sections are known with subpercent precision~\cite{Vogel:1999zy,Kurylov:2001av,Kurylov:2002vj,Strumia:2003zx,Ankowski:2016oyj,Ricciardi:2022pru,Tomalak:2025okl,Tomalak:2025jtn}, which matches the accuracy goals of reactor antineutrino experiments such as Jiangmen Underground Neutrino Observatory (JUNO). Should a supernova explosion occur in our galaxy, precise IBD cross sections would be crucial for analyzing the (anti)neutrino signals observed at Earth. A detailed investigation of IBD cross sections started more than 40 years ago with the first evaluations of quantum electrodynamics (QED) radiative corrections in Refs.~\cite{Vogel:1983hi,Fayans:1985uej}, slightly after the formulation of the current algebra approach to radiative corrections~\cite{Sirlin:1977sv}. In the current century, radiative corrections to IBD have been reformulated using modern effective field theories (EFTs). First, bottom-up EFT for QED radiative corrections in neutron decay was introduced in~\cite{Ando:2004rk}. Then, this group suggested the same framework of pionless heavy-baryon chiral perturbation theory (HBChPT) for IBD~\cite{Raha:2011aa}. The matching to theory with pions was performed in Ref.~\cite{Cirigliano:2022hob}. However, the relevant low-energy coupling constants (LECs), vector $g_V$ and axial-vector $g_A$, were not fully determined. Subsequently, a framework for the determination of LECs was developed and applied for the vector coupling constant in Refs.~\cite{Cirigliano:2023fnz,Cirigliano:2024nfi} by incorporating all higher-energy electroweak and quantum chromodynamics (QCD) physics. With a precise theoretical prediction for $g_V$, the QED radiative corrections to IBD were comprehensively evaluated in pionless HBChPT in Refs.~\cite{Tomalak:2025okl,Tomalak:2025jtn}. The latter references worked at reactor antineutrino energies $E_{\overline{\nu}_e} \lesssim 10~\mathrm{MeV}$ when pion contributions can be safely neglected. In this work, we aim to extend QED radiative corrections in IBD to higher energies by including kinematic-dependent corrections from pion degrees of freedom.

The matching of pionless HBChPT for charged-current processes with nucleons to the theory with pions revealed potentially large few-percent contributions to the axial-vector charge $g_A$~\cite{Cirigliano:2022hob,Tomalak:2026wks} not only from the leading order (LO) but also from the Wilson coefficient of the next-to-leading order (NLO) HBChPT $c_4-c_3$~\cite{Hoferichter:2015tha,Hoferichter:2015hva,Hall:2025ytt,Tomalak:2026wks} and confirmed the poor convergence of the HBChPT expansion~\cite{Hoferichter:2015tha,Hoferichter:2015hva}. To determine whether IBD is subject to potential enhancements from NLO HBChPT and to investigate its convergence pattern, we calculate the QED radiative corrections to IBD within the HBChPT framework, including pion degrees of freedom.

The paper is organized as follows. In Section~\ref{sec:IBD}, we describe kinematics and leading-order cross sections in the IBD reaction. We evaluate corrections from virtual pion fields in Section~\ref{sec:QED}. First, we discuss QED radiative corrections in pionless effective field theory. Subsequently, we describe contributions from the one-pion exchange diagram in subsection~\ref{subsec:pion_exchange}. In subsection~\ref{subsec:LO_HBChPT}, we present the leading-order QED radiative corrections with virtual pions. Corrections from the next-to-leading order HBChPT Lagrangian are provided in subsection~\ref{subsec:NLO_HBChPT}. In Section~\ref{sec:results_and_discussion}, we present numerical results for the kinematic dependence of pion-induced contributions. Our conclusions and outlook are collected in Section~\ref{sec:conclusions_and_outlook}. The Wolfram Mathematica notebook and a Python library for fast and accurate evaluation of IBD cross sections are available at~\href{https://github.com/tomalak7/IBDxsec}{github.com/tomalak7/IBDxsec}.

\section{Inverse beta decay} \label{sec:IBD}

We introduce IBD kinematics and discuss the allowed kinematic region in Section~\ref{subsec:kinematics}. In Section~\ref{subsec:leading_order}, we describe IBD and present the corresponding cross section at leading order.

\subsection{Kinematics in inverse beta decay} \label{subsec:kinematics}

\begin{figure}[ht]
\begin{center}
	\includegraphics[scale=1.]{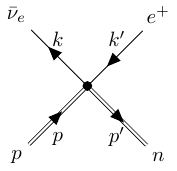}
	\caption{Kinematics in inverse beta decay (IBD). \label{fig:diagram}}
\end{center}
\end{figure}

We illustrate the inverse beta decay reaction $\overline{\nu}_e p \to e^+ n$ and denote the momenta of the incoming and outgoing particles in figure~\ref{fig:diagram}.
The initial proton is taken at rest: $p^\mu = (m_p,~0)$, with the proton mass $m_p$. The final neutron momentum satisfies $p^{\prime2} = m_n^2$, with the neutron mass $m_n$. The incoming antineutrino with momentum $k^\mu = (E_{\overline{\nu}_e},~\bk)$ can be described as a massless particle: $k^2 = 0$. The final positron has the momentum $k^{\prime\mu} = (E_e,~\bk^\prime)$: $k^{\prime2} = m_e^2$ and the mass $m_e$. We denote the momentum transfer as $q^\mu = k^\mu - k^{\prime\mu}$. The Lorentz-invariant squared momentum transfer $Q^2$ can be expressed in terms of the masses and energies of the particles as
\begin{equation} \label{eq:squared_momentum_transfer}
	Q^2 = - q^2 = -\left(p^\prime - p \right)^2 = m^2_p - m^2_n + 2 m_p \left( E_{\overline{\nu}_e} - E_e \right).
\end{equation}
For convenience, we also introduce the squared energy in the center-of-mass reference frame $s$,
\begin{equation} \label{eq:squared_cmf_energy}
	s = \left(p + k \right)^2 = m^2_p + 2 m_p E_{\overline{\nu}_e}.
\end{equation}

The kinematically allowed recoil positron energy $E_e$ in IBD lies in the narrow energy range $E_e^\mathrm{min} \le E_e \le E_e^\mathrm{max}$:
\begin{align}
	E_e^\mathrm{min} &= \frac{ \left( s - m^2_n + m^2_e \right) \left( m_p + E_{\overline{\nu}_e} \right) - E_{\overline{\nu}_e} \sqrt{\Sigma \left( s, m^2_n, m^2_e \right)} }{2 s}, \label{eq:positron_energy_min} \\
	E_e^\mathrm{max} &= \frac{ \left( s - m^2_n + m^2_e \right) \left( m_p + E_{\overline{\nu}_e} \right) + E_{\overline{\nu}_e} \sqrt{\Sigma \left( s, m^2_n, m^2_e \right)} }{2 s}, \label{eq:positron_energy_max}
\end{align}
with the kinematic triangle function $\Sigma \left( s, m^2_n, m^2_e \right) = \left( s - \left( m_n + m_e \right)^2 \right) \left( s - \left( m_n - m_e \right)^2 \right)$. This function is positive-definite above the inverse beta decay threshold $E_{\overline{\nu}_e} \ge E_{\overline{\nu}_e}^\mathrm{thr} = \frac{\left( m_n + m_e \right)^2}{2 m_p} - \frac{m_p}{2}$, with $E_{\overline{\nu}_e}^\mathrm{thr} \approx 1.806066~\mathrm{MeV}$, while charged-current elastic neutrino-neutron scattering is a thresholdless process.

\subsection{IBD at leading order} \label{subsec:leading_order}

To evaluate IBD cross sections, we start with the effective Lagrangian $\mathcal L$~\cite{Ando:2004rk,Falkowski:2021vdg,Cirigliano:2022hob,Tomalak:2023xgm,Cirigliano:2023fnz,Cirigliano:2024nfi} of low-energy charged-current lepton-nucleon interactions:
\begin{equation} \label{eq:Lagrangian_at_leading_order}
	\mathcal L = - \sqrt{2} G_F V^\star_{ud} \overline{\nu}_{eL} \gamma_\rho e \cdot \overline{N}_v \left( g_V v^\rho - 2 g_A S^\rho \right) \tau^- N_v + \mathrm{h.c.},
\end{equation}
where QCD and electroweak contributions from energy scales above the pion mass are included in the LECs: the vector $g_V$ and the axial-vector $g_A$. $G_F = 1.1663787(6)\times 10^{-5}~\mathrm{GeV}^{-2}$~\cite{Fermi:1934hr,Feynman:1958ty,vanRitbergen:1999fi,MuLan:2012sih} is the scale-independent Fermi coupling constant, $V_{ud} = 0.97348(31)$~\cite{Cabibbo:1963yz,Kobayashi:1973fv,ParticleDataGroup:2020ssz,Hardy:2020qwl} is the Cabibbo-Kobayashi-Maskawa quark mixing matrix element. $v^\rho = (1,0)$ denotes the nucleon velocity, $S^\rho$ is the nucleon spin: $v \cdot S = 0$. The heavy-nucleon fields are combined in the doublet $N_v = \left( p, n \right)^T$, with the proton $p$ and neutron $n$ fields, respectively. The raising and lowering operators in the isospin space are expressed in terms of the Pauli matrices $\tau^1$ and $\tau^2$ as $\tau^\pm = \frac{\tau^1 \pm i \tau^2 }{2}$. The positron and antineutrino are described by the fields $e$ and ${\nu}_{eL}$, respectively.

After straightforward evaluation of the spin-averaged squared matrix element from the contact interaction in Eq.~(\ref{eq:Lagrangian_at_leading_order}), the leading-order differential IBD cross section can be written as
\begin{equation} \label{eq:LO_static_diff}
	\frac{\mathrm{d}\sigma_{\rm LO}}{\mathrm{d} Q^2} = \frac{{G}_{F}^2 |V_{ud}|^2}{2\pi} \bigg[ \left( 1 - \frac{E_0}{E_{\overline{\nu}_e}} \right) \left( g_V^2 + g_A^2 \right) - \frac{Q^2+m^2_e}{4 E_{\overline{\nu}_e}^2} \left( g_V^2 - g_A^2 \right) \bigg],
\end{equation}
with $E_0 = \frac{ m^2_n + m^2_e - m^2_p}{2 m_n} \approx 1.292581~\mathrm{MeV}$, which can be approximated as the neutron-proton mass difference: $E_0 \approx m_n - m_p$.

\section{QED radiative corrections in IBD} \label{sec:QED}

\begin{figure}[ht]
	\centering
	\includegraphics[scale=1.]{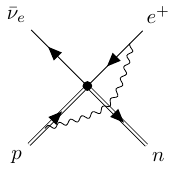}
	\caption{One-loop virtual QED correction in IBD. \label{fig:one_loop_QED}}
\end{figure}

In pionless effective field theory, QED radiative corrections are formulated and evaluated by extending the low-energy Lagrangian of Eq.~(\ref{eq:Lagrangian_at_leading_order}) with QED interactions. Besides the renormalization of external charged particles, only one virtual diagram with the exchange of the photon between the positron and the proton contributes at leading order in $\alpha$ expansion, cf. figure~\ref{fig:one_loop_QED}.

\begin{figure}[ht]
	\centering
	\includegraphics[scale=1.]{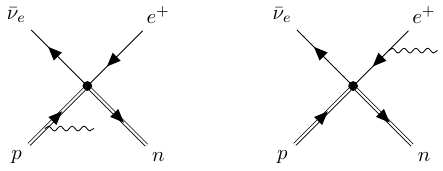}
	\caption{Leading in nucleon recoil one-photon bremsstrahlung contributions to radiative IBD process: $\overline{\nu}_e + p \rightarrow e^+ + n + \gamma$. \label{fig:bremsstrahlung_graphs}}
\end{figure}

To properly describe the long-distance contributions and provide cross-section predictions with a percent level of accuracy, the bremsstrahlung from the positron and the proton, shown in figure~\ref{fig:bremsstrahlung_graphs}, also have to be included. The detailed derivation and analytical expressions for these contributions are presented in Refs.~\cite{Tomalak:2025jtn,Tomalak:2025okl}.

In this work, we extend these calculations by including virtual pion fields. For convenience, we directly exploit the results of Refs.~\cite{Tomalak:2025jtn,Tomalak:2025okl} with the renormalized LECs $g_V$ and $g_A$ in Eq.~(\ref{eq:Lagrangian_at_leading_order}) and consider only the kinematic dependence of the pion-induced contributions. 

We study chiral interactions of pion, nucleon, photon, and lepton fields. To incorporate pion degrees of freedom $\vec{\pi}$ entering $u$, $u^2 = U = e^{\frac{i \boldsymbol{\pi} \cdot \boldsymbol\tau}{F_\pi}}$,  with the pion decay constant $F_\pi$, we start with the HBChPT Lagrangian at lowest order~\cite{Gasser:1983yg,Meissner:1997ii,Muller:1999ww,Knecht:1999ag,Gasser:2002am}
\begin{equation}
    \mathcal L^{p^2}_\pi + \mathcal L^{e^2}_\pi + \mathcal L^{p}_{\pi N} = \frac{F_\pi^2}{4} \langle u_\mu u^\mu + m^2_\pi \left( U^\dagger + U \right) \rangle  + e^2 Z_\pi F_\pi^4 \langle U  Q U^\dagger Q  \rangle +  \overline{N}_v i v \cdot \nabla N_v + g^{(0)}_A \overline{N}_v S \cdot u N_v, \label{eq:LO}
\end{equation}
where $g_A^{(0)}$ is the nucleon axial-vector coupling constant in the chiral limit, $Z_\pi$ is the QED isospin-breaking LEC responsible for the electromagnetic pion-mass splitting, and $m_\pi$ is the pion mass. The trace of any object ${\cal  O}$ in the isospin space is denoted as $\langle {\cal  O} \rangle$. The standard ChPT building blocks are expressed in terms of the external left and right sources $l_\mu$ and $r_\mu$, respectively, as
\begin{align}
    u_\mu &=  i \left[ u^\dagger (\partial_\mu - i r_\mu) u  - u (\partial_\mu - i l_\mu) u^\dagger\right], \\
    \nabla_\mu N_v &\equiv  \left(\partial_\mu + \Gamma_\mu  \right) N_v,  \qquad \Gamma_\mu = \frac{1}{2} \left[ u (\partial_\mu - i l_\mu) u^\dagger + u^\dagger (\partial_\mu - i r_\mu) u \right].
\end{align}
To evaluate low-energy charged-current processes with leptons and nucleons, we account for the electromagnetic and electroweak interactions through the external sources as
\begin{align}
    l_\mu &= - e Q A_\mu  + Q_W \, \overline{e} \gamma_\mu \nu_{eL} + Q_W^\dagger \overline{\nu}_{eL} \gamma_\mu e, \\
    r_\mu &= - e Q A_\mu, \label{eq:sources}
\end{align}
with the charge matrix $Q = {\rm diag} (1, 0)$ and the electroweak matrix $Q_W = -2 \sqrt{2} {G}_{F} V_{ud} \tau^+$.

To account for the corrections suppressed by $\frac{m_\pi}{4 \pi F_\pi} \sim \frac{m_\pi}{m_n}$, we consider the $\mathcal O(p^2)$ and $\mathcal O(e^2)$ pion-nucleon interaction Lagrangians~\cite{Gasser:1987rb,Krause:1990xc,Ecker:1995rk,Bernard:1995dp,Meissner:1997ii,Muller:1999ww}\footnote{For consistency with the notations of Refs.~\cite{Tomalak:2025jtn,Tomalak:2025okl}, we exploit the neutron mass $m_n$ for recoil corrections.}
\begin{align}
&  \mathcal L^{p^2}_{\pi N} + \mathcal L^{e^2}_{\pi N} =\overline{N}_v \Bigg[ \frac{(v \cdot \mathcal \nabla)^2 - \mathcal \nabla^2}{2 m_n}  - \frac{i g^{(0)}_A}{2 m_n} \left\{ S \cdot {\nabla}, v \cdot u\right\} + c_1 m^2_\pi \langle U^\dagger + U \rangle \nonumber \\
&+  \left(c_2 - \frac{\left( g^{\left(0 \right)}_A\right)^2 }{8 m_n}\right) (v\cdot u)^2 + c_3 u \cdot u + \left(c_4 + \frac{1}{4 m_n} \right) \left[S^\mu, S^\nu\right] u_\mu u_\nu  + c_5 m^2_\pi \left( \tilde{U}^\dagger + \tilde{U} \right) \nonumber \\
&- \frac{i \left[S^\mu,S^\nu\right]}{4 m_n}   \left( (1+\kappa_1) f^+_{\mu\nu} + \frac{1}{2} (\kappa_0 - \kappa_1) \langle  f^+_{\mu\nu}\rangle \right) \Bigg] N_v \nonumber \\
& + e^2 F^2_\pi \, \overline{N}_v \left( f_1   \langle \tilde{{\cal  Q}}_+^2 - {\cal  Q}_-^2 \rangle + f_2 \tilde{{\cal  Q}}_+  \langle {\cal  Q}_+ \rangle +  f_3  \langle \tilde{{\cal  Q}}_+^2 + {\cal  Q}_-^2 \rangle + f_4 \langle {\cal  Q}_+ \rangle^2 + f_5 {\cal  Q}_-  \langle {\cal  Q}_+ \rangle  \right) N_v, \label{eq:HBChPT_nlo}
\end{align}
with $\tilde{{\cal  O}} = {\cal  O} - \frac{1}{2} \langle {\cal  O} \rangle$, $f^+_{\mu\nu} =  u^\dagger \left(  \partial_\mu r_\nu -  \partial_\nu r_\mu - i \left[ r_\mu, r_\nu \right] \right) u + u \left(  \partial_\mu l_\nu -  \partial_\nu l_\mu - i \left[ l_\mu, l_\nu \right] \right) u^\dagger$, and the matrices ${{\cal{Q}}}_\pm$:
\begin{equation}
	{{\cal{Q}}}_\pm = \frac{u Q u^\dagger \pm u^\dagger Q u}{2}.
\end{equation}
The LECs $c_1-c_5$, $f_1-f_5$ are under active investigation~\cite{Hoferichter:2015tha,Hoferichter:2015hva,Siemens:2016jwj,Hall:2025ytt,Tomalak:2026wks}, $\kappa_{0}$ and $\kappa_{1}$ represent the nucleon isoscalar and isovector anomalous magnetic moments, respectively.

\subsection{Pion exchange diagram} \label{subsec:pion_exchange}

\begin{figure}[ht]
\begin{center}
	\includegraphics[scale=1.0]{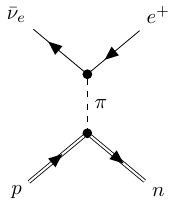}
	\caption{One-pion exchange diagram. \label{fig:one_pion_exchange_diagram}}
\end{center}
\end{figure}

The pion exchange diagram contributes to charged-current semi-leptonic processes involving nucleons at tree level. We illustrate the diagram in figure~\ref{fig:one_pion_exchange_diagram}.
The one-pion exchange generates the matrix element $T^{1\pi}$,
\begin{equation}
    T^{1\pi} =  \frac{2 m_e g_A^{(0)}}{q^2 - m^2_\pi} \sqrt{2}  G_F V^\star_{ud} \overline{\nu}_{e L} e \cdot \overline{N}_v S \cdot q \tau^- N_v ,
\end{equation}
suppressed by the electron mass. The corresponding cross-section contributions from the squared matrix element and from the interference terms are suppressed by $2$ powers of the electron mass and, therefore, can be neglected.

\subsection{Leading-order HBChPT contributions} \label{subsec:LO_HBChPT}

\begin{figure}[ht]
\begin{center}
	\includegraphics[scale=0.615]{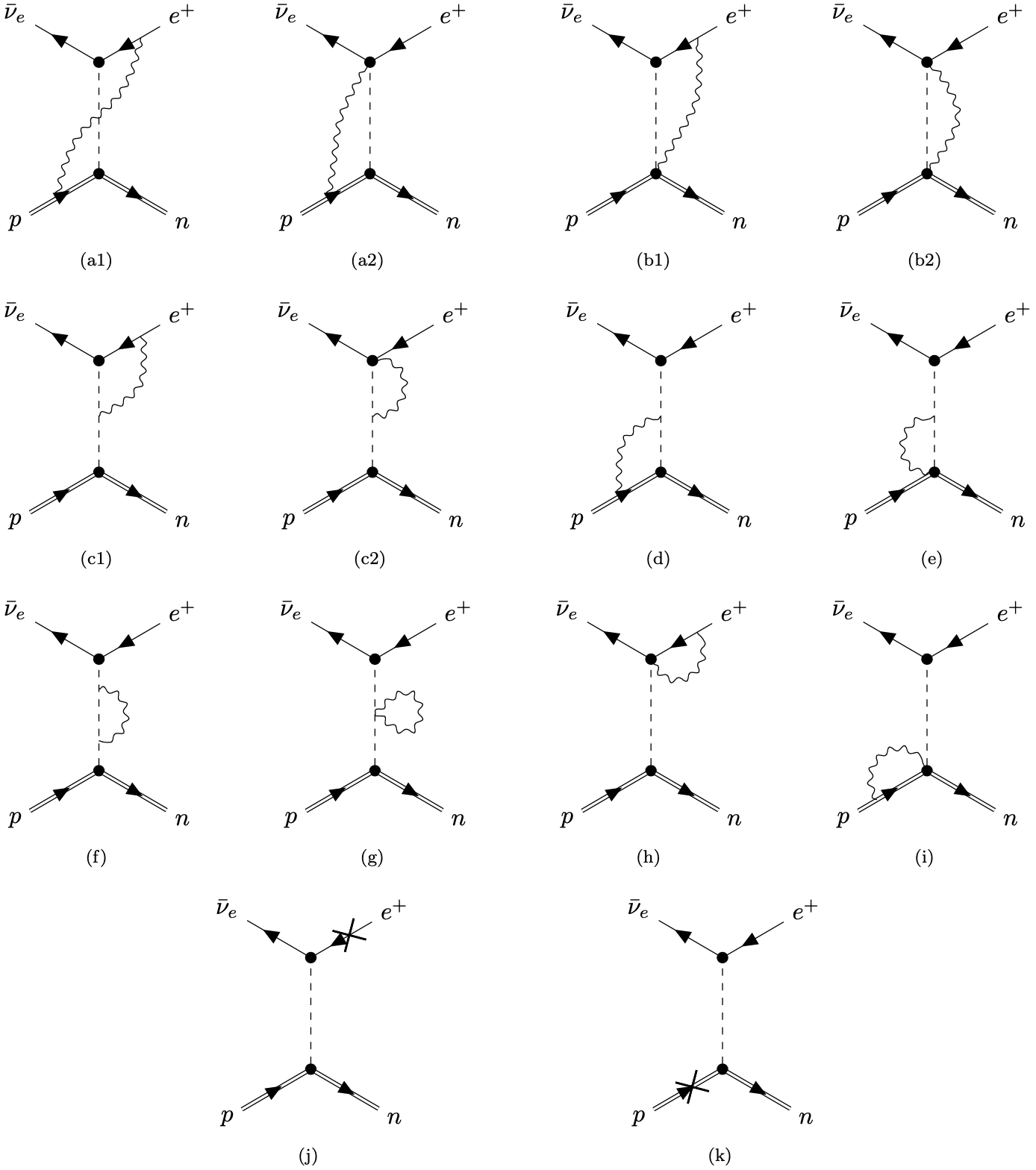}
	\caption{QED radiative corrections with virtual photons at leading order. The field renormalization contributions (j) and (k) to the one-pion exchange diagram in Section~\ref{subsec:pion_exchange} are also shown. \label{fig:leading_order_diagrams_with_photons}}
\end{center}
\end{figure}

Photon couplings to the positron, the proton, and the pion at leading HBChPT order generate diagrams in figure~\ref{fig:leading_order_diagrams_with_photons}. The sum of all these corrections is gauge-invariant.

Considering the sum of any of two diagrams in figure~\ref{fig:leading_order_diagrams_with_photons} with a photon of the momentum $L$ attached to the positron line and to the charged-current leptonic vertex, we obtain the following expression from the lepton line in the momentum space:
\begin{equation} \label{eq:fermion_simplifucation}
	1 +  \frac{\left(\slashed{L} - \slashed{q} \right) \left( - \slashed{p}_e - \slashed{L} + m_e  \right) }{ L^2 + 2 p_e \cdot L } \to - \frac{m_e^2}{ L^2 + 2 p_e \cdot L } + \frac{ \left(\hat{L} +  \hat{p}_e  \right) m_e}{ L^2 + 2 p_e \cdot L } = \mathcal{O} \left( m_e \right).
\end{equation}
Consequently, the following diagrams in figure~\ref{fig:leading_order_diagrams_with_photons} are suppressed by the electron mass in pairs: (a1) and (a2), (b1) and (b2), (c1) and (c2), while all other diagrams without the coupling of the photon to the positron are also suppressed by the electron mass similar to the one-pion exchange in Section~\ref{subsec:pion_exchange}. Therefore, we can neglect the contribution from all diagrams in figure~\ref{fig:leading_order_diagrams_with_photons}, both in forward and non-forward kinematics. Moreover, the same argument applies to the pion-induced bremsstrahlung at the amplitude level. This simplification is equivalent to using equations of motion in the tree-level Lagrangian~\cite{Tomalak:2021lif}.

\begin{figure}[ht]
\begin{center}
	\includegraphics[scale=0.345]{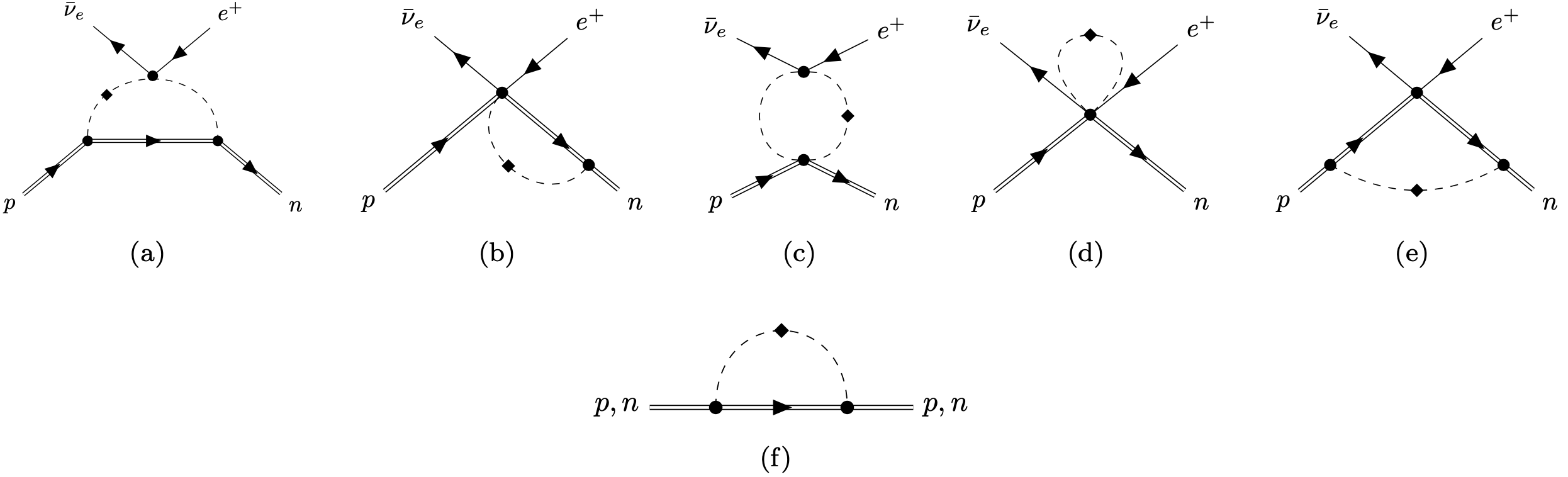}
	\caption{QED radiative corrections without virtual photons at leading order. The pion isospin-breaking vertex is shown as a square, and we imply all possible placements of this vertex. The nucleon field-renormalization diagram (f) is also illustrated. \label{fig:leading_order_Zpi_diagrams}}
\end{center}
\end{figure}

The hard-photon contributions to the charged pion mass are included in the LEC $Z_\pi$ of the Lagrangian in Eq.~(\ref{eq:LO}). This isospin-breaking interaction contributes to the proton and neutron field renormalization factors and generates the diagrams in figure~\ref{fig:leading_order_Zpi_diagrams}.
All contributions from the pion isospin-breaking interaction are manifestly gauge-invariant and do not generate any bremsstrahlung at the order of our calculation. Consequently, all the radiation from the leading-order HBChPT interactions is suppressed by the electron mass and does not require precise calculations. 

In the following, we provide the results for each diagram in figure~\ref{fig:leading_order_Zpi_diagrams}, neglecting the terms suppressed by the electron mass. We use dimensional regularization in $d = 4 - 2 \varepsilon$ with the chiral version of modified minimal subtraction scheme~\cite{Gasser:1983yg} and subtract the following expression with the ultraviolet pole $\frac{1}{\varepsilon}$
\begin{equation}
    \frac{1}{\varepsilon} - \gamma_E + \ln \left( 4 \pi \right) + 1.
\end{equation}
We validate master integrals from Refs.~\cite{Cho:1992cf,Falk:1993fr,Boyd:1994pa,Stewart:1998ke,Bouzas:1999ug,Fajfer:2001ad,Bouzas:2001py,Davydychev:2001ui,Zupan:2002je,Bouzas:2002xi,Becirevic:2002sc} in this calculation. However, we were unable to validate the analytical expressions for the four-point master integrals in Ref.~\cite{Zupan:2002je}, which do not enter our final results, in either nonforward or forward kinematics.

a) The contribution of this diagram can be represented as the kinematic-dependent change of the vector coupling constant:
\begin{equation} \label{eq:LO_agV}
	\delta g_V = - \frac{\alpha}{4\pi}  Z_\pi \left( g_A^{(0)} \right)^2 \left[ 3 \left( \ln \frac{\mu^2_\chi}{m^2_\pi} - 1 \right) - 2 \right] - \frac{\alpha}{4\pi}  Z_\pi \left( g_A^{(0)} \right)^2 \left[ 4 + \left( 2 \beta + \frac{1}{\beta}  \right) \ln \frac{\beta - 1}{\beta + 1} \right],
\end{equation}
with $\beta = \sqrt{1+\frac{4m^2_\pi}{Q^2}}$, and a non-factorizable correction to the IBD cross section
\begin{equation} \label{eq:LO_c_NF}
	\frac{\mathrm{d}\sigma}{\mathrm{d} Q^2} = - \alpha Z_\pi \left( g_A^{(0)} \right)^2 \arcsin{\beta^{-1}} \frac{{G}_{F}^2 |V_{ud}|^2}{2\pi} \frac{Q}{E_{\overline{\nu}_e}} g_A.
\end{equation}

b) This diagram does not contribute to charged-current semi-leptonic processes with a nucleon.

c) The contribution of this diagram can be represented as the kinematic-dependent change of the vector coupling constant:
\begin{equation} \label{eq:LO_c}
	\delta g_V = - \frac{\alpha}{4\pi}  Z_\pi \left( \ln \frac{\mu^2_\chi}{m^2_\pi} - 1 \right) - \frac{\alpha}{4\pi}  Z_\pi \left( 2 + \beta \ln \frac{\beta - 1}{\beta + 1} \right).
\end{equation}

d) The tadpole diagram does not depend on the kinematics of external particles and contributes to the renormalization of the vector and axial-vector LECs as
\begin{equation} \label{eq:LO_d}
	\delta g_V = \frac{\delta g_A}{g_A^{(0)}} = \frac{\alpha}{4\pi}  Z_\pi \left( \ln \frac{\mu^2_\chi}{m^2_\pi} - 1 \right).
\end{equation}

e) This diagram does not contribute to charged-current semi-leptonic processes with the nucleon.

f) The nucleon field renormalization factors renormalize the vector and axial-vector LECs as
\begin{equation} \label{eq:Zp_LO}
	\delta g_V = \frac{\delta g_A}{g_A^{(0)}} = \frac{\alpha}{4 \pi} Z_\pi \left( g_A^{(0)} \right)^2 \left[ 3 \left( \ln \frac{\mu^2}{m^2_\pi} - 1 \right) - 2 \right].
\end{equation}

Combining all pieces together, we reproduce the renormalization of the vector and axial-vector LECs at leading order~\cite{Cirigliano:2022hob}
\begin{align} \label{eq:renormalization_LO}
	\delta g^\mathrm{LO}_V &= 0, \nonumber \\
    \frac{\delta g^\mathrm{LO}_A}{g_A^{(0)}} &= \frac{\alpha}{2 \pi} Z_\pi \left[  \frac{ 1 + 3 \left( g_A^{(0)} \right)^2}{2} \left( \ln \frac{\mu^2}{m^2_\pi} - 1 \right) -  \left( g_A^{(0)} \right)^2 \right].
\end{align}
These contributions are included as part of the LECs $g_V$ and $g_A$ in the calculation of the effective field theory without pions. The leftover kinematic dependence of the LO pion-induced QED radiative corrections can be represented as the shift in the vector coupling constant,
\begin{equation} \label{eq:gV_shift_LO}
	g_V \to g_V - \frac{\alpha}{2 \pi}  Z_\pi \left[ \frac{ 1 + 2 \left( g_A^{(0)} \right)^2}{2}\left(2 + \beta \ln \frac{\beta - 1}{\beta + 1} \right) + \frac{\left( g_A^{(0)} \right)^2}{2 \beta}  \ln \frac{\beta - 1}{\beta + 1} \right],
\end{equation}
and additional non-factorizable cross section $\mathrm{d}\sigma^{\rm NF}_{\rm LO}$,
\begin{equation} \label{eq:LO_NF}
	\frac{\mathrm{d}\sigma^{\rm NF}_{\rm LO}}{\mathrm{d} Q^2} = - \alpha Z_\pi \left( g_A^{(0)} \right)^2 \arcsin{\beta^{-1}} \frac{{G}_{F}^2 |V_{ud}|^2}{2\pi} \frac{Q}{E_{\overline{\nu}_e}} g_A.
\end{equation}
The correction $\mathrm{d}\sigma^{\rm NF}_{\rm LO}$ is the dominant kinematic-dependent pion-induced QED contribution that scales as $\alpha \frac{E_{\overline{\nu}_e}}{m_\pi}$. The subleading correction in Eq.~(\ref{eq:gV_shift_LO}) enters as $\frac{\alpha}{\pi} \frac{E^2_{\overline{\nu}_e}}{m^2_\pi}$.

\subsection{Next-to-leading-order HBChPT contributions} \label{subsec:NLO_HBChPT}

At next-to-leading order, the same topologies as in figures~\ref{fig:one_pion_exchange_diagram} and~\ref{fig:leading_order_diagrams_with_photons} contribute, but with NLO instead of LO HBChPT vertices on the nucleon side. The contribution from the sum of these diagrams to virtual and real radiative corrections is suppressed by the electron mass and, therefore, can be neglected.

\begin{figure}[ht]
\begin{center}
	\includegraphics[scale=1.]{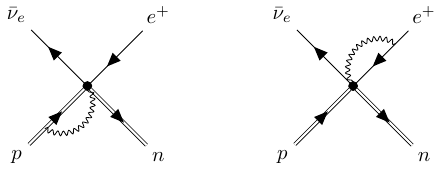}
	\caption{QED radiative corrections from interactions in Eq.~(\ref{eq:recoil_lagrangian_with_photon}). \label{fig:recoil_diagrams_with_photons_only}}
\end{center}
\end{figure}

At order $\mathcal{O} \left( \frac{1}{m_n} \right)$, an additional contact interaction with a radiation of one photon is present. The corresponding Lagrangian $\mathcal{L}^\gamma_\mathrm{recoil}$ can be expressed as 
\begin{align} \label{eq:recoil_lagrangian_with_photon} 
	\mathcal{L}^\gamma_\mathrm{recoil} &= \frac{e}{2 m_n} \sqrt{2} G_F V^\star_{ud} \left[ \overline{\nu}_{e L} \slashed{A} e \cdot \overline{N}_v  \tau^- N_v - \overline{\nu}_{e L} \slashed{v} A_\rho e \cdot \overline{N}_v \left( v^\rho - 2 S^\rho  \right) \tau^- N_v \right] \nonumber \\
    &+  \frac{e}{m_n} \left( 1 + \kappa_1 \right) \sqrt{2} G_F V^\star_{ud}  \overline{\nu}_{e L} \gamma_\rho e \cdot \overline{N}_v \left[ S \cdot A, S^\rho \right] \tau^- N_v.  
\end{align}
This interaction generates virtual contributions in IBD by exchanging photons with either the positron or the proton. We illustrate the corresponding Feynman diagrams in figure~\ref{fig:recoil_diagrams_with_photons_only}.
These contributions are gauge-invariant and infrared-finite. The diagram with a coupling to the proton vanishes. Nevertheless, the corresponding corrections are not suppressed by the electron mass; they are small compared to the leading contribution from the diagram (a) in figure~\ref{fig:leading_order_Zpi_diagrams}. These corrections can be estimated as $\frac{\alpha}{\pi}\frac{E_{\overline{\nu}_e}}{m_n}$, resulting in effects well below $0.1\%$. The radiation of real photons that enters as an interference with the leading-order radiation is expected to have an effect of a similar size on the IBD cross section and, therefore, can be neglected.

At next-to-leading order, the nucleon magnetic moment contributes radiative-recoil corrections with a topology of the diagram in figure~\ref{fig:one_loop_QED} and a diagram with electromagnetic coupling to the neutron instead of the proton. Moreover, the recoil contributions to the contact interaction in figure~\ref{fig:diagram} enter with the field renormalization factors and with an exchanged photon in the topology of one-loop correction in figure~\ref{fig:one_loop_QED}. These contributions do not involve virtual pion fields and, therefore, can be estimated as $\frac{\alpha}{\pi}\frac{E_{\overline{\nu}_e}}{m_n}$, similar to the analysis of diagrams in figure~\ref{fig:recoil_diagrams_with_photons_only}. We neglect the corresponding corrections and consider only the $Z_\pi$-dependent virtual pion-induced contributions that are not suppressed by the electron mass.

At next-to-leading order, the same topologies as in figure~\ref{fig:leading_order_Zpi_diagrams} contribute, but with recoil corrections and NLO instead of LO HBChPT vertices on the nucleon side. Neglecting the renormalization of the nucleon magnetic moments and higher-order operators, as well as of the electron-mass suppressed terms, we provide the contributions from each topology in figure~\ref{fig:leading_order_Zpi_diagrams} in the following.

a) The correction from this diagram, removing the renormalization of the nucleon magnetic moments, and higher-order operators, can be represented as the kinematic-dependent change of the vector coupling constant:
\begin{equation} \label{eq:NLO_a_gV}
	\delta g_V = - \frac{15}{8} \alpha Z_\pi \left( g_A^{(0)} \right)^2 \frac{m_\pi}{m_n} + \alpha Z_\pi \left( g_A^{(0)} \right)^2 \left( 1 + \frac{1}{4 \beta^2} - \frac{\beta^2 \arcsin{\beta^{-1}}}{\sqrt{\beta^2 -1}}\right) \frac{m_\pi}{2 m_n},
\end{equation}
and non-factorizable recoil correction,
\begin{align} \label{eq:NLO_a_NF}
	\frac{\mathrm{d}\sigma}{\mathrm{d} Q^2} &= \frac{\alpha}{\pi} Z_\pi \left( g_A^{(0)} \right)^2 \left[ 11 \left( 2 + \beta \ln \frac{\beta - 1}{\beta + 1} \right) + \frac{1 + 3 \beta^2}{3} + \frac{\beta^4-5}{2 \beta} \ln \frac{\beta - 1}{\beta + 1} \right]  \nonumber \\
    &\times \frac{{G}_{F}^2 |V_{ud}|^2}{2\pi} \frac{Q^2}{12 m_n E_{\overline{\nu}_e}} g_A.
\end{align}

b) The contribution of this diagram can be represented as the kinematic-dependent change of the vector coupling constant:
\begin{equation} \label{eq:NLO_b_gV}
	\delta g_V = \frac{3}{4} \alpha Z_\pi \left( g_A^{(0)} \right)^2 \frac{m_\pi}{m_n} + \alpha Z_\pi \left( g_A^{(0)} \right)^2 \frac{Q^2}{8 m_\pi m_n},
\end{equation}
and the renormalization of the axial-vector coupling constant:
\begin{equation} \label{eq:NLO_b_gA}
	\frac{\delta g_A}{g_A^{(0)}} = 2 \alpha Z_\pi m_\pi \left[ c_4 - c_3 + \frac{3}{8 m_n} \right].
\end{equation}

c) This diagram results in a non-factorizable NLO correction from the Wilson coefficient $c_4$:
\begin{equation} \label{eq:NLO_NF_c4}
	\frac{\mathrm{d}\sigma}{\mathrm{d} Q^2} = - \frac{\alpha}{\pi} Z_\pi \left( c_4 + \frac{1}{4 m_n} \right)  \left( 2 + \beta \ln \frac{\beta - 1}{\beta + 1} \right) \frac{{G}_{F}^2 |V_{ud}|^2}{2\pi} \frac{Q^2}{E_{\overline{\nu}_e}} g_A.
\end{equation}

d) Besides the renormalization of the nucleon magnetic moments, this diagram does not contribute to charged-current semi-leptonic processes with nucleons at the order of our calculation.

e) This diagram does not contribute to charged-current semi-leptonic processes with a nucleon.

f) NLO HBChPT interactions from Eq.~(\ref{eq:HBChPT_nlo}) contribute to the proton and neutron field renormalization factors as a recoil correction,
\begin{equation} \label{eq:Zp_NLO}
	\delta g_V = \frac{\delta g_A}{g_A^{(0)}} =  \frac{9}{8} \alpha Z_\pi \left( g_A^{(0)} \right)^2\frac{m_\pi}{m_n}.
\end{equation}
Results for the nucleon field renormalization factors can be readily applied to all low-energy processes involving the nucleon.

Combining all pieces together, we reproduce the renormalization of the vector and axial-vector coupling constants at next-to-leading order~\cite{Cirigliano:2022hob}
\begin{align} \label{eq:renormalization_NLO}
	\delta g^\mathrm{NLO}_V &= 0, \\
    \frac{\delta g^\mathrm{NLO}_A}{g_A^{(0)}} &= 2 \alpha Z_\pi m_\pi \left[ c_4 - c_3 + \frac{3}{8 m_n} + \frac{9}{16} \frac{\left( g_A^{\left( 0 \right)} \right)^2}{m_n} \right].
\end{align}
These contributions are included as part of the LECs $g_V$ and $g_A$ in the calculation of the effective field theory without pions. The remaining kinematic dependence of the NLO pion-induced QED radiative corrections can be represented as the shift in the vector coupling constant
\begin{equation} \label{eq:gV_shift_NLO}
	g_V \to g_V + \alpha Z_\pi \left( g_A^{(0)} \right)^2 \left( 1 + \frac{1}{4 \beta^2} - \frac{\beta^2 \arcsin{\beta^{-1}}}{\sqrt{\beta^2 -1}}\right) \frac{m_\pi}{2 m_n} + \alpha Z_\pi \left( g_A^{(0)} \right)^2 \frac{Q^2}{8 m_\pi m_n},
\end{equation}
and an additional non-factorizable cross section
\begin{align} \label{eq:NLO_NF}
	\frac{\mathrm{d}\sigma^{\rm NF}_{\rm NLO}}{\mathrm{d} Q^2} = &- \frac{\alpha}{\pi} Z_\pi \left( c_4 + \frac{1}{4 m_n} - \frac{11}{12} \frac{\left( g_A^{\left( 0 \right)} \right)^2}{m_n} \right)  \left( 2 + \beta \ln \frac{\beta - 1}{\beta + 1} \right) \frac{{G}_{F}^2 |V_{ud}|^2}{2\pi} \frac{Q^2}{E_{\overline{\nu}_e}} g_A \nonumber \\
    &+\frac{\alpha}{\pi} Z_\pi \left( g_A^{(0)} \right)^2 \left[ \frac{1 + 3 \beta^2}{3} + \frac{\beta^4-5}{2 \beta} \ln \frac{\beta - 1}{\beta + 1} \right] \frac{{G}_{F}^2 |V_{ud}|^2}{2\pi} \frac{Q^2}{12 m_n E_{\overline{\nu}_e}} g_A.
\end{align}
Only the Wilson coefficient $c_4$ contributes to the kinematic dependence of the pion-induced QED radiative corrections at next-to-leading order.

\section{Results and Discussion} \label{sec:results_and_discussion}

In this Section, we present our results for the kinematic-dependent pion-induced QED radiative corrections at LO and NLO HBChPT in Eqs.~(\ref{eq:gV_shift_LO}),~(\ref{eq:LO_NF}),~(\ref{eq:gV_shift_NLO}), and~(\ref{eq:NLO_NF}).

At antineutrino energy below the pion mass, we find the leading contribution $\sim \alpha \frac{E_{\overline{\nu}}}{m_\pi}$ from the diagram (a) in figure~\ref{fig:leading_order_Zpi_diagrams} with leading-order HBChPT couplings. The diagrams in figure~\ref{fig:leading_order_Zpi_diagrams} also contribute terms $\lesssim \alpha \frac{m_\pi}{m_n}$ and $\lesssim \frac{\alpha}{\pi}  \frac{E^2_{\overline{\nu}_e}}{m^2_\pi}$. We fully include them in our results, accounting also for higher-order terms in the kinematic expansion. The diagrams in figure~\ref{fig:recoil_diagrams_with_photons_only} and the corresponding real radiation enter as $\sim \frac{\alpha}{\pi} \frac{E_{\overline{\nu}_e}}{m_n}$ and, therefore, are suppressed compared to the leading contribution by a factor $\sim \frac{m_\pi}{\pi m_n}$. Therefore, we can safely neglect the pionless NLO diagrams. To account for potentially enhanced contributions from the NLO Wilson coefficients, we include NLO contributions with the topology of diagrams in figure~\ref{fig:leading_order_Zpi_diagrams}. Moreover, we neglect all radiative contributions that scale as the electron mass $\lesssim \alpha \frac{m_e}{m_\pi}$ from figure~\ref{fig:leading_order_Zpi_diagrams} and $\lesssim \frac{\alpha}{\pi}  \frac{m_e}{E_{\overline{\nu}_e}}$ from figure~\ref{fig:leading_order_diagrams_with_photons} and neglect the leading-order one-pion exchange diagram from figure~\ref{fig:one_pion_exchange_diagram} that enters as $\lesssim\frac{m^2_e}{m^2_\pi}$.

\begin{figure}[ht]
\begin{center}
	\includegraphics[scale=0.35]{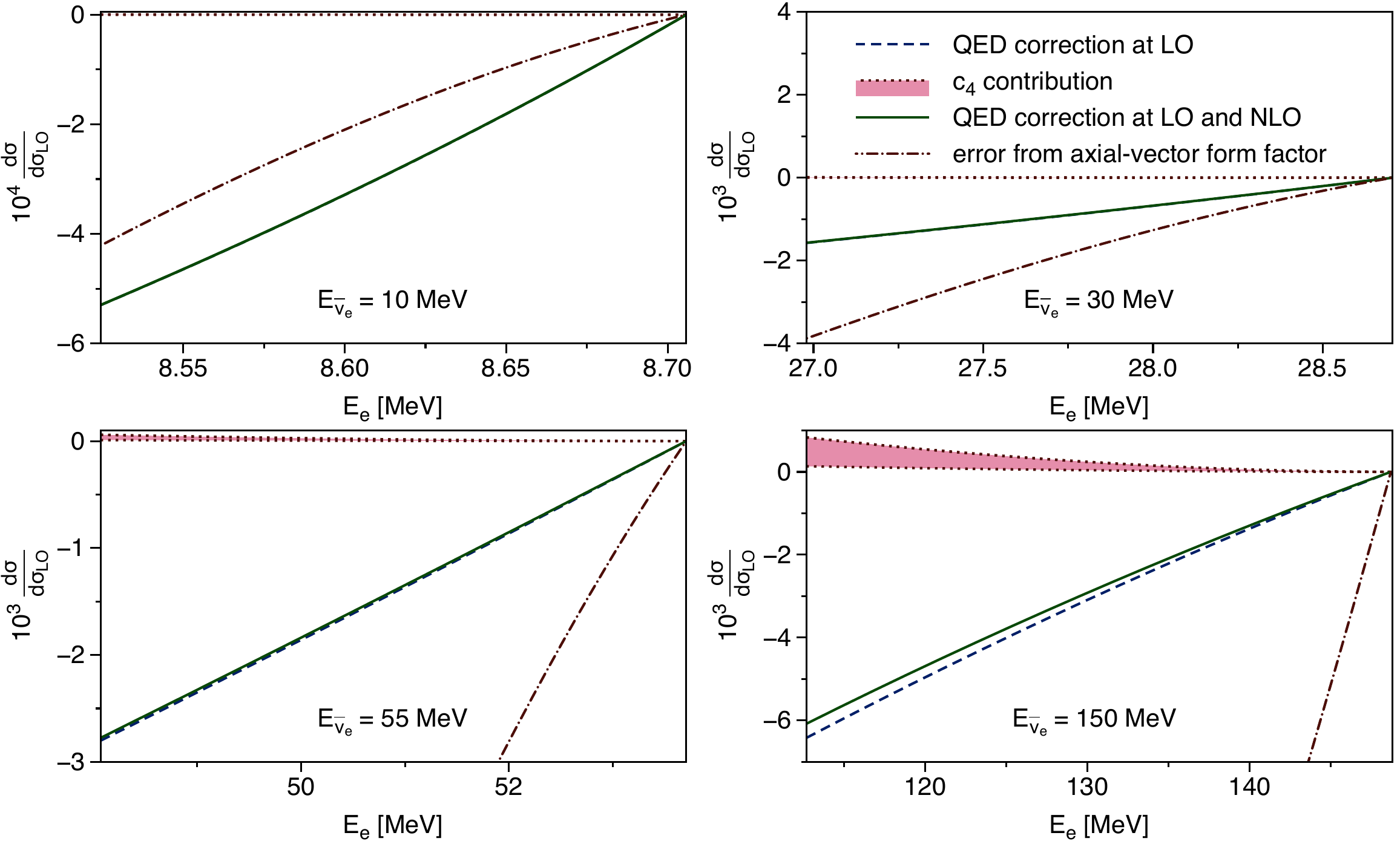}
	\caption{Ratio of the kinematic-dependent pion-induced QED radiative contributions $\sigma$ to the leading order cross section $\sigma_{\mathrm{LO}}$ is shown as a function of the recoil positron energy $E_e$ for antineutrino beam energies $E_{\overline{\nu}_e} = 10, 30, 55,$ and $150~\mathrm{MeV}$. The leading-order contribution, shown as the blue dashed line, is compared to the contribution from the Wilson coefficient $c_4$, shown as the red dotted line and a pink band corresponding to the range of inputs from Refs.~\cite{Hoferichter:2015tha,Hoferichter:2015hva,Hall:2025ytt,Tomalak:2026wks}, to the total kinematic-dependent pion-induced QED correction, shown by the green solid line, and to the taken with an opposite sign uncertainty from the nucleon axial-vector form factor~\cite{MINERvA:2023avz,Tomalak:2026wsu}, with fixed normalization, shown by the red dash-dotted line. \label{fig:pion_induced_corrections}}
\end{center}
\end{figure}

We show kinematic-dependent QED radiative corrections to IBD cross sections in figure~\ref{fig:pion_induced_corrections} as a function of the recoil positron energy for antineutrino beam energies $E_{\overline{\nu}_e} = 10, 30, 55,$ and $150~\mathrm{MeV}$.
We compare the resulting correction, the sum of LO and NLO contributions, with the LO HBChPT result and find an expected convergence pattern, with the dominant contribution from LO. We investigate the dependence of radiative corrections on the NLO HBChPT LECs. Only the Wilson coefficient $c_4$ contributes to the kinematic dependence of pion-induced QED radiative corrections. Substituting the range of its values from Refs.~\cite{Hoferichter:2015tha,Hoferichter:2015hva,Hall:2025ytt,Tomalak:2026wks}, we find negligibly small effects in comparison with LO radiative corrections. Even at energies $E_{\overline{\nu}_e} \sim 150~\mathrm{MeV}$, the contribution from $c_4$ is below $0.1\%$. However, this correction reaches $0.16\%$ at $E_{\overline{\nu}_e} \sim 200~\mathrm{MeV}$. Consequently, IBD cross sections with pion decay at rest neutrinos and supernova (anti)neutrinos are not sensitive to the poor convergence of the HBChPT expansion.

\begin{figure}[]
\begin{center}
	\includegraphics[scale=0.5]{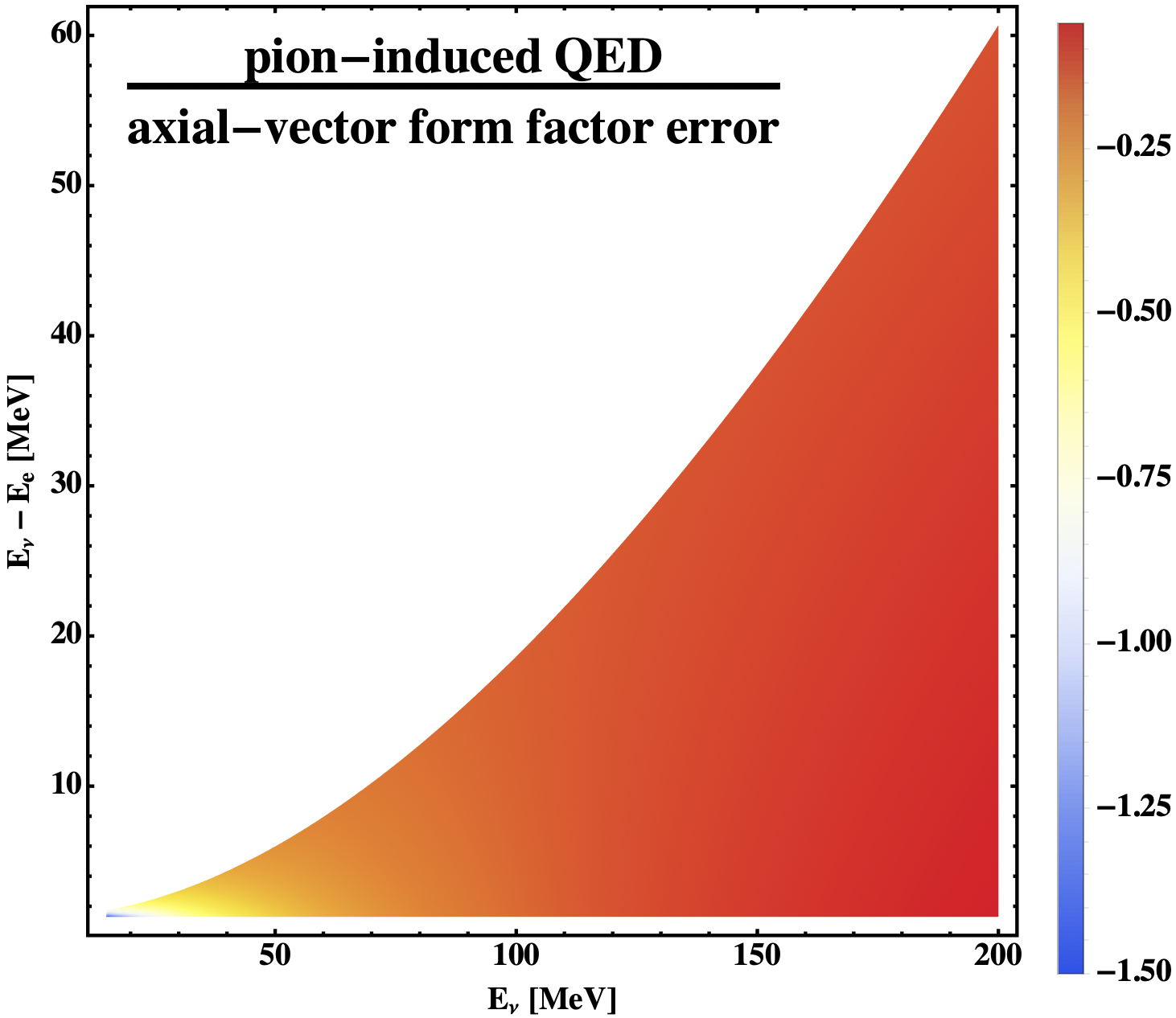}
	\caption{Ratio of the kinematic-dependent pion-induced QED radiative contributions to the uncertainty from the nucleon axial-vector from factor, with fixed normalization, is shown for the kinematically allowed range of positron energy when the antineutrino energy $15~\mathrm{MeV} \le E_{\overline{\nu}_e} \le 200~\mathrm{MeV}$. \label{fig:axial_vector_form_factor_error}}
\end{center}
\end{figure}

In figure~\ref{fig:axial_vector_form_factor_error}, we illustrate the size of the kinematic-dependent pion-induced contributions to IBD in comparison to the uncertainty from the nucleon axial-vector form factor~\cite{Meyer:2016oeg,MINERvA:2023avz,Tomalak:2026wsu}, with fixed normalization. Currently, the contribution from pion loops is at the level and below our knowledge of the nucleon form factors. Only for energies $E_{\overline{\nu}_e} \lesssim 15~\mathrm{MeV}$, the kinematic-dependent pion-induced QED corrections are above the form-factor errors. However, the results of this work will become feasible for pion decay at rest and supernova (anti)neutrinos after new precise measurements of the momentum dependence of the nucleon axial-vector form factor with improved precision~\cite{Petti:2023abz}.

The pion-induced QED radiative corrections in the charged-current neutrino-neutron elastic scattering and neutron decay have the same form and can be obtained by interchanging the proton and neutron masses, $m_p \leftrightarrow m_n$.

At $E_\nu \sim m_\pi$ and above, on top of all corrections in the pionless theory~\cite{Tomalak:2025jtn,Tomalak:2025okl}, the diagrams with an intermediate $\Delta(1232)$ resonance must be systematically included to obtain a complete prediction for the (differential) cross section~\cite{Jenkins:1990jv,Hemmert:1996xg,Hemmert:1996rw,Hemmert:1997ye,Kambor:1997ns,Pascalutsa:2002pi,Geng:2008mf,Krebs:2009bf}.

\section{Conclusions and Outlook} \label{sec:conclusions_and_outlook}

In this paper, we perform the first calculation of pion-induced QED radiative corrections in inverse beta decay. We account for all leading- and next-to-leading order contributions within the framework of the heavy baryon chiral perturbation theory. Accounting for pion degrees of freedom renormalizes the nucleon axial-vector coupling constant at a few $\%$ level. However, these large radiative corrections are fully incorporated when the experimental value of the nucleon axial-vector charge is exploited in the pionless calculation. We find the dominant leading-order kinematic-dependent contribution from pions reaching a few $0.1\%$ level at antineutrino energies $E_{\overline{\nu}_e} \gtrsim 20~\mathrm{MeV}$. Contrary to the renormalization of the nucleon axial-vector charge, the kinematic dependence of next-to-leading order pion-induced radiative contributions in IBD, which is governed by the Wilson coefficient $c_4$ and the recoil corrections, is subdominant and might become relevant only at antineutrino energies above the pion mass. Consequently, the poor HBChPT convergence does not affect the phenomenology of reactor antineutrino as well as pion decay at rest and supernova (anti)neutrino physics at the single-nucleon level. We find contributions from virtual pions to the kinematic dependence of IBD cross sections below the uncertainty from the momentum dependence of the nucleon axial-vector form factor for energies $E_{\overline{\nu}_e} \gtrsim 15~\mathrm{MeV}$. This work enables predictions of the inverse beta decay cross sections at antineutrino energies $E_{\overline{\nu}_e} \gtrsim 10~\mathrm{MeV}$ with $0.1\%$ precision and below, assuming ideal inputs in terms of the momentum dependence of the nucleon form factors, the quark mixing matrix, and the nucleon axial-vector charge.

\acknowledgments

We thank Zhewen Mo for his help in preparing the figures and for his insightful comments. We also thank Zhewen Mo and Jiang-Hao Yu for useful discussions. This work is supported by the National Science Foundation of China under Grants No. 12347105 and No. 12447101. FeynCalc~\cite{Mertig:1990an,Hahn:1998yk,Shtabovenko:2016sxi}, Mathematica~\cite{Mathematica}, and DataGraph~\cite{JSSv047s02} were used in this work.

\bibliographystyle{JHEP}
\bibliography{references}

\end{document}